\documentclass[twocolumn,letter]{jpsj2}
\title{
Isostaticity in two dimensional pile of rigid disks}

\author{Akihiro \textsc{Kasahara} and
 Hiizu \textsc{Nakanishi}\thanks{%
E-mail address: naka4scp@mbox.nc.kyushu-u.ac.jp
}}

\inst{
Department of Physics, Kyushu University 33, Fukuoka 812-8581
}

\abst{ We study the static structure of piles made of polydisperse disks
in the rigid limit with and without friction using molecular
dynamic simulations for various elasticities of the disks and pile
preparation procedures.  The coordination numbers are calculated to
examine the isostaticity of the pile structure.  For the frictionless
pile, it is demonstrated that the coordination number converges to 4 in
the rigid limit, which implies that the structure of rigid disk pile
is isostatic. On the other hand, for the frictional case with the
infinite friction constant, the coordination number depends on the
preparation procedure of the pile, but we find that the structure
becomes very close to isostatic with the coordination number close to 3
in the rigid limit when the pile is formed through the process that
tends to make a pile of random configuration.  }

\kword{
static structure of granular pile, isostaticity, marginal rigidity,
stress propagation
}

\begin{document}
\maketitle


Isostaticity is a peculiar property that a rigid granular pile is
expected to have\cite{M98}.  Consider first a pile of frictionless deformable
grains under gravity.  The pile is assumed to be stable against small
enough external perturbation.  Each grain in the pile should be deformed
and the forces acting through the contacts between the grains are
elastic force, that is related to the grain deformation through
elasticity.  Imagine that the grains in the pile are being hardened
gradually.  As the grains become harder, their deformation diminishes
and some of the contacts disappear.  In the rigid limit, no deformation
is allowed and the forces acting between the grains should be determined
only by the structure of contact network of the pile without any
information of deformation.

Based on such observation, it has been conjectured\cite{M98} that the
contact network in a static pile of non-cohesive rigid grains without
friction should have a property that the network has just enough
contacts to be rigid in the sense that removal of any contact makes the
structure flexible.  This property is called {\em
isostaticity}\cite{M98,M01} or {\em marginal rigidity}\cite{BB02,BB03}:
the structure should be unstable against an infinitesimal small external
force if there are less contacts than this, while it should be
overconstrained with internal stress if there are more contacts.

It is not obvious, however, whether this holds true or not for the pile
that is formed through deposition of grains that are rigid from the
beginning.  It is even less obvious for the case that the friction
between the grains comes into the problem.


A simple way to check whether a pile is isostatic is just to count the
average coordination number $z$ of grains in the pile.  Since all the
contact forces in the isostatic pile have to be determined without
redundant constraint, the number of independent components of the forces
should coincide with the number of the conditions for the force balance
and the torque balance for each grain.  This leads to $z=2d$ for a pile
of frictionless spheres, $z=d(d+1)$ for a pile of frictionless
non-spherical grains, and $z=d+1$ for a frictional pile of both
spherical and non-spherical grains in $d$ dimensions.

Isostaticity for the piles deposited from a certain initial
configuration has been examined by numerical simulations for three
dimensional monodisperse spherical systems with various elastic
constants\cite{SEGHL02}, and it was shown that the piles of frictionless
sphere becomes isostatic in the hard sphere limit, while the pile of
frictional spheres are more dependent on the way how the piles are
prepared, but are not likely to be isostatic.

On the other hand, Blumenfeld et al. performed simple tabletop
experiments for two dimensional non-spherical grains of cardboard and
concluded that the piles of two dimensional grains become isostatic
in the limit that the number density of a initial configuration and that
of a final stable configuration coincides.


In this report, using numerical simulations for two dimensional system,
we examine the piles of two dimensional polydisperse disk system for
several ways of preparation procedures.

\begin{figure}[b]
\begin{center}
\includegraphics[height=3cm,angle=-90]{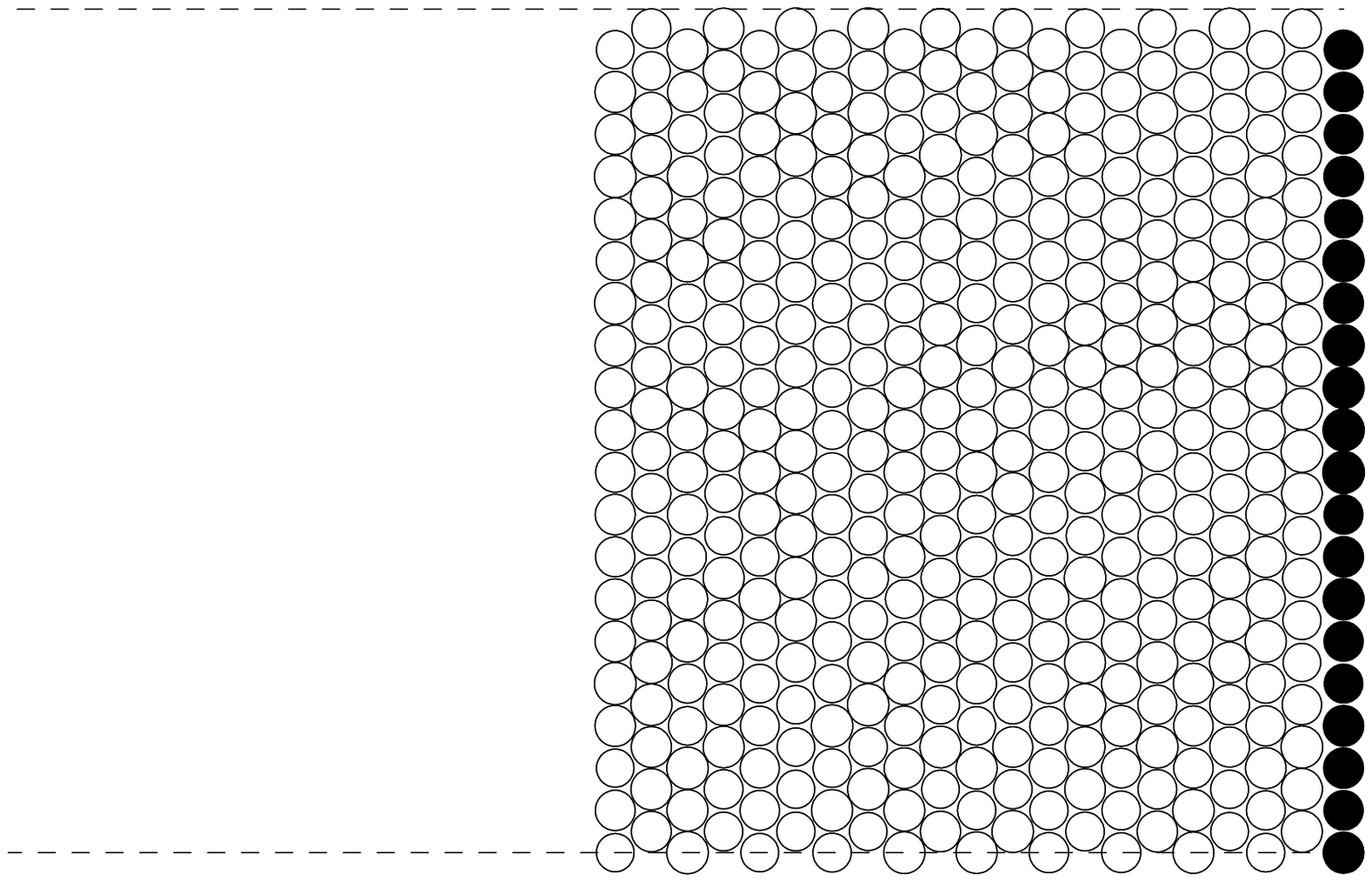}\hspace{1cm}
\includegraphics[height=3cm,angle=-90]{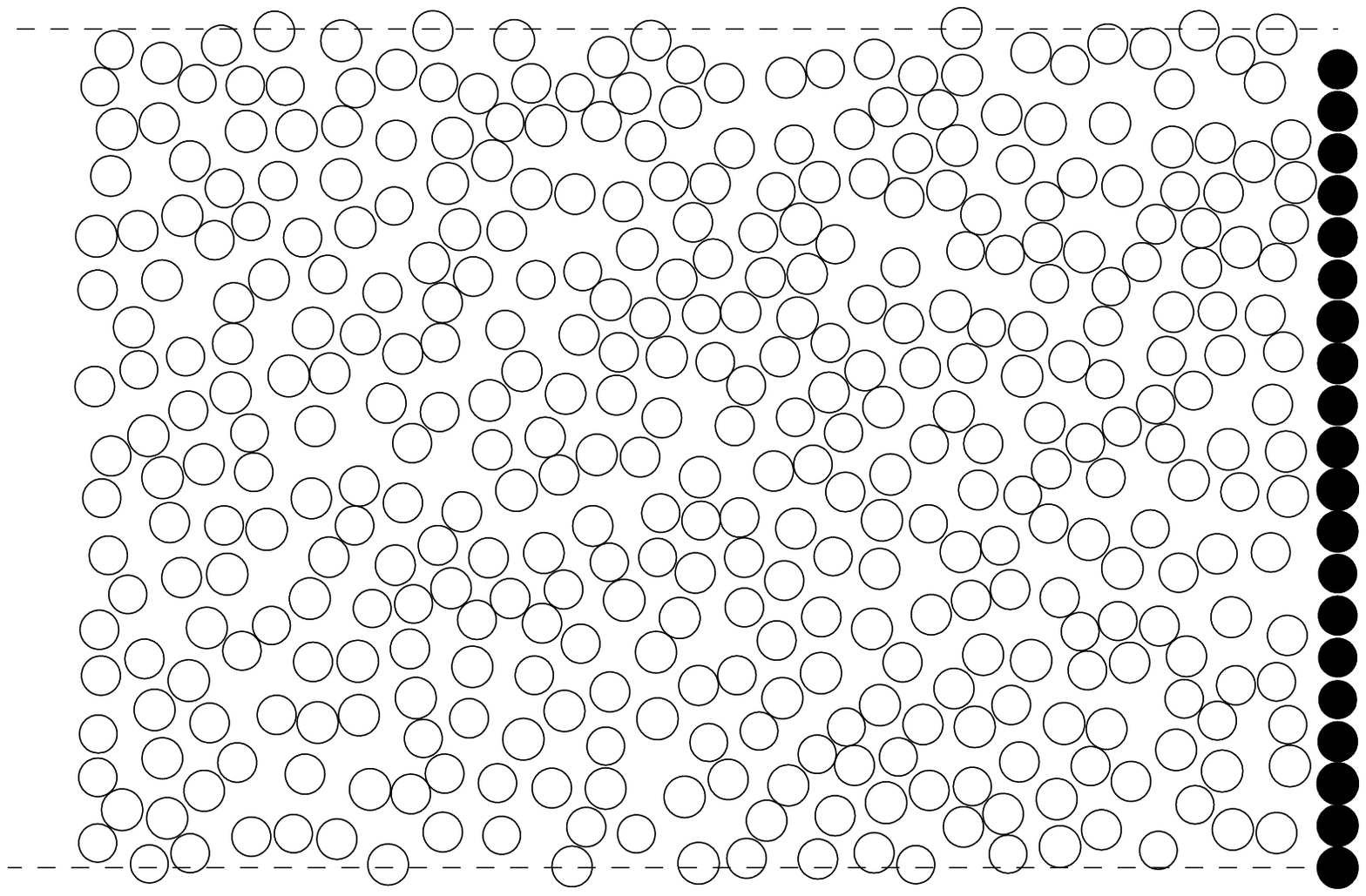}
\end{center}
\caption{
Two initial configurations:  the triangular lattice with polydisperse
 disks (left) and the random configuration with the number density 0.6
 (right).
}
\label{f-1}
\end{figure}

We performed molecular dynamics simulations on the system that consists
of two dimensional disks with linear elasticity and damping.  The system
is polydisperse with the uniform distribution in the disk diameter with
the range $[0.9,1]$ of the maximum diameter $\sigma_0$, and their masses
are assumed to be proportional to their areas: the mass of the disk with
the diameter $\sigma_0$ is denoted by $m_0$.  The bottom of the system
is made rough by attaching the disks with the interval $\sigma_0$, and
we employ the periodic boundary condition in the horizontal direction.
The number of the disks $N$ in the system is 400, and the horizontal
length of the system is $20\sigma_0$, thus we have approximately 20
layers of disks on average.

Piles are formed by letting the system run under the gravitational
acceleration $g$ from initial configurations until all the disks stop
moving.

\begin{figure}[tb]
\begin{center}
\includegraphics[width=8cm]{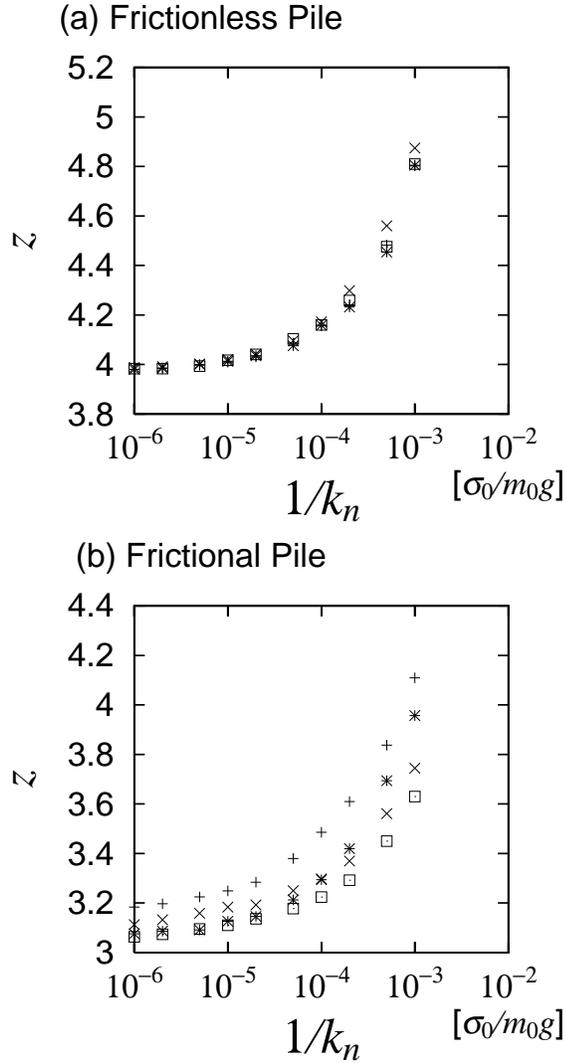}
\end{center}
\caption{ The coordination number $z$ for various elastic constant $k_n$
for frictionless pile (a) and frictional pile (b).  The marks represent
the pile preparation procedure: the Newtonian equation with the
triangular lattice initial configuration ($+$), the Newtonian equation
with the random initial configuration ($\times$), the viscous equation
with the triangular lattice initial configuration ($*$), the viscous
equation with the random initial configuration ($\square$).  Each mark
represents average over six to twelve realizations.  } \label{f-2}
\end{figure}
\begin{figure}[tb]
\begin{center}
\includegraphics[width=8cm]{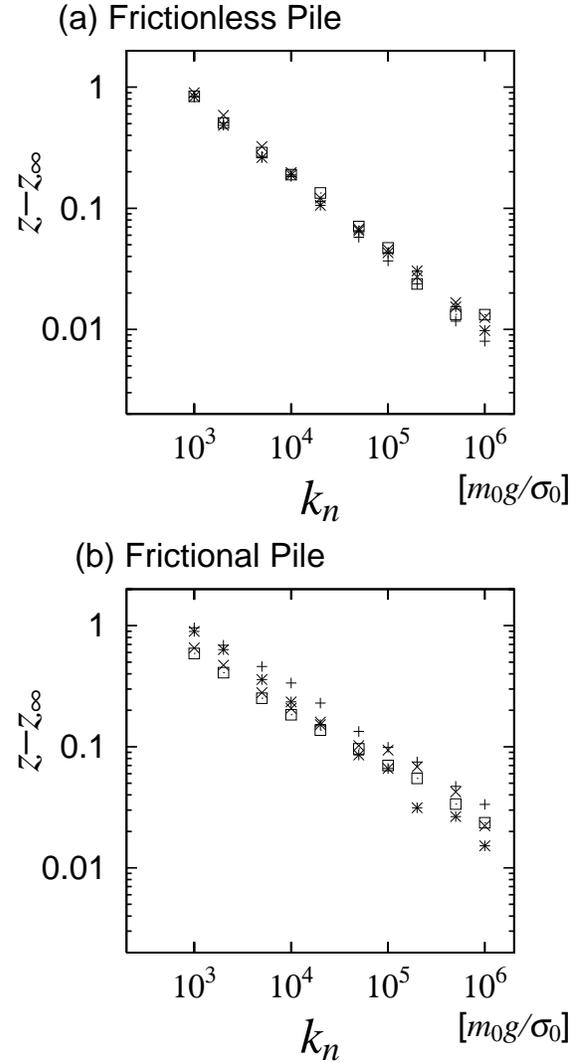}
\end{center}
\caption{ The coordination numbers $z$ for various elastic constant
$k_n$ in the log-log scale frictionless pile (a) and frictional pile
(b).  The same data are plotted using the same marks with those in
Fig.\ref{f-2} } \label{f-3}
\end{figure}

The disk at the position $\mib{x}(t)$ with mass $m$ follows the
Newtonian equation
\begin{equation}
m\ddot{\mib{x}}(t)=\mib{F}(t),
\label{Newton}
\end{equation}
where the force $\mib{F}(t)$ consists of the gravitation and the contact
force from the neighboring disks in contact.  The two particles
$i$ and $j$ at $\mib{x}_i(t)$ and $\mib{x}_j(t)$ with radii $r_i$ and
$r_j$, respectively, are in contact when the overlap $\delta_{ij}$ given by
\begin{equation}
\delta_{ij}=r_i+r_j-|\mib{x}_i-\mib{x}_j|,
\label{delta}
\end{equation}
is positive.  Then the particle $j$ exerts the force $\mib{F}_{ij}$  on
the particle $i$
\begin{equation}
\mib{F}_{ij} = \mib{F}_{ij}^n  +\mib{F}_{ij}^t ,
\end{equation}
where $\mib{F}_{ij}^n$ and $\mib{F}_{ij}^t$ are the normal and
tangential components of the force;
\begin{eqnarray}
\mib{F}_{ij}^n & = & k_n\delta_{ij}\mib{\hat{n}}_{ij}-\gamma_n \mib{v}_{ij}^n ,
\label{normal}
\\
\mib{F}_{ij}^t & = & -k_t\Delta s_{ij}\mib{\hat{t}}_{ij}-
   \gamma_t \mib{v}_{ij}^t .
\label{tangential}
\end{eqnarray}
Here, $\mib{\hat n}_{ij}$ and $\mib{\hat t}_{ij}$ are the normal and tangential unit
vectors, respectively;
 $\Delta s_{ij}$ is the tangential displacement of the contact
points after the contact, $\mib{v}_{ij}^n$ ($\mib{v}_{ij}^t$) is
relative normal (tangential) velocity, $k_n$ ($k_t$) is the normal
(tangential) elastic constant, and $\gamma_n$ ($\gamma_t$) is the normal
(tangential) damping constant.  
Note that we assume no threshold for
 the disks to slip during the contact, which corresponds to
the case with the infinite friction coefficient.
In the case of frictionless disk, we
simply set $\mib{F}_{ij}^t=0$.


In the actual simulations, we use $k_t=0.2 k_n$ and
$\gamma_n=2\sqrt{k_n} \,[\sqrt{m_0}]$ for the frictionless case, and
$k_t=0.2 k_n$ and $\gamma_n=\gamma_t=2\sqrt{k_n} \,[\sqrt{m_0}]$ for
the frictional case.

As for the initial configurations, we try the triangular lattice and the
random configuration (Fig.\ref{f-1}); the triangular lattice with the
lattice constant $\sigma_0$ is not a regular lattice because the disks
located at the lattice points are polydisperse.  The random
configurations are prepared by putting disks at random with the area
fraction around 0.6.

Simulations to form piles start from these configurations with zero
particle velocity and stop when the kinetic energy of each disks
becomes negligibly small, namely smaller than $10^{-15}[m_0g\sigma_0]$.


Fig.\ref{f-2}(a) shows the $k_n$-dependence of the coordination number $z$ for
the frictionless disks.  It can be seen that the preparation dependence
is very small and $z$ converges to the number very close to 4 in the
large $k_n$ limit for both of the initial configurations.  The
$k_n$-dependence is well represented by the power law
\begin{equation}
z-z_\infty \propto k_n^{-\alpha}
\end{equation}
as is shown in Fig.\ref{f-3}(a).

As for the case of frictional disks, the results are shown in
Fig. \ref{f-2}(b).  There are two things to be noted in comparison with
the frictionless case: (i) the discrepancy between the two initial
conditions shown by $+$ and $\times$ is large, (ii) the limiting values of the
coordination number is substantially different from 3, or the value for
the isostatic structure of the frictional grain in two dimensions.

The limiting values of the coordination numbers are $z_\infty=3.15$ for
the triangular lattice initial condition and $z_\infty=3.09$ for the
random initial condition(Table \ref{t-1}).
\begin{table}[tb]
\begin{tabular}{|cc|cc|cc|}
\hline
Eq. of motion & Initial config.  & 
\multicolumn{2}{c|}{Frictionless} &
\multicolumn{2}{c|}{Frictional} \\
& & $z_\infty$ & $\alpha$ & $z_\infty$ & $\alpha$ 
 \\
\hline
Newtonian & Triangular lattice & 3.97  & 0.68  & 3.15 & 0.49
\\
 & Random config. & 3.98 & 0.65 & 3.09 & 0.47
\\ \hline
Viscous & Triangular lattice & 3.97 & 0.63 & 3.06 & 0.60
\\
 & Random config. & 3.97 & 0.64 & 3.04 & 0.46
\\
\hline
\end{tabular}
\caption{
The limiting coordination numbers $z_\infty$ and the exponents $\alpha$
 for various preparation procedures.
}
\label{t-1}
\end{table}

These features indicate that the piles of frictional disks are more
dependent on the way they are formed than those of the frictionless
piles.  The fact that $z_\infty$ for the random initial conditions is
closer to 3 suggests that the piles would tend to be isostatic when they
are formed through the process that retains a random structure.

This consideration prompts us to employ the viscous equation
\begin{equation}
\gamma\dot{\mib{x}}(t) = \mib{F}(t)
\label{viscous}
\end{equation}
for the time development during the deposition process in stead of the
Newtonian equation, in order to see if it makes the structure closer to
the isostaticity.  The viscous equation tends to produce random
configurations in equilibrium because the disks stop as soon as the
force balance is achieved without the inertia effect.  In the simulation
with the viscous equation, we take $\gamma=5[m_0\sqrt{g/\sigma_0}]$ with
$\gamma_n=\gamma_t=0$ in eqs.(\ref{normal}) and (\ref{tangential}).

The data for the piles by the viscous equation are also shown in
Fig.\ref{f-3} and Table \ref{t-1}.
One can see that the limiting values for the viscous equation are
actually closer to 3 than those for the Newtonian equation.


In summary, in this report we have shown the followings; The piles of
frictionless disks become isostatic when the disks are very hard and
they are not sensitive to the preparation procedure very much, which is
consistent with the conjecture that the pile of rigid grains is
isostatic.  On the other hand, for the piles of frictional disks with
the infinite friction, the structure depends on the preparation process.
If the pile is formed from the triangular lattice with the inertia, the
pile structure seems to be distinctively different from the isostatic
one even in the rigid limit, as has been found in the previous work on
the three dimensional system\cite{SEGHL02}.  We found, however, that
the pile of frictional disks becomes very close to isostatic in the
rigid grain limit when we employ the deposition process that tends to
produce random structure.

The isostaticity of granular pile should play an crucial role in the
stress propagation of granular system.  In the case of frictionless
pile, it has been shown that the isostatic structure has some peculiar
features: (i) unidirectional nature of stress propagation\cite{TW99},
(ii) correspondence between the force-force response and the
displacement-displacement response\cite{M01}, (iii) unstable response
that grows as the distance from the point of an external
perturbation\cite{M98}.  The effects of friction on the mechanical
properties of the isostatic structure are now under investigation.


\end{document}